\documentclass[twocolumn,showpacs,preprintnumbers,amsmath,amssymb,aps,prb,superscriptaddress,floatfix]{revtex4}
\usepackage{graphicx}
\usepackage[latin1]{inputenc}

\begin{document}
\title{Spin nematic phases in models of correlated electron systems: a numerical study}
\author{S.\ Capponi} \affiliation{Laboratoire de
Physique Th\'eorique, CNRS UMR 5152, Universit\'e Paul Sabatier, F-31062 Toulouse, France.}
\author{F.~F.~Assaad} \affiliation{Institut f\"ur Theoretische Physik und Astrophysik,
Universit\"at W\"urzburg, Am Hubland, D-97074 W\"urzburg, Germany}

\date{\today}

\begin{abstract}
Strongly interacting systems are known to often 
spontaneously develop exotic ground states under certain
conditions. For instance, spin nematic phases have been discovered in various magnetic models. Such 
phases, which break spin symmetry but have no net local magnetization, have also been proposed by Nersesyan {\it et al.}~\cite{Nersesyan91} in the
context of electronic models. We introduce a $N$-flavor microscopic model that interpolates from the large-$N$ limit, where mean-field is valid and 
such a nematic phase occurs, to the more
realistic $N=1$ case. By using a sign-free quantum Monte-Carlo, we show the existence of a spin nematic phase (analogous to a  spin flux phase) for
finite $N$; when $N$ decreases, quantum fluctuations increase and this phase ultimately disappears in favor of an s-wave superconducting state. We also show that 
 this nematic phase extends up to a finite critical charge doping. 

Dynamical studies allow us to clarify the Fermi surface property~: in the nematic phase at half-filling, it consists of four points and the low-energy 
structure has a Dirac cone-like shape. Under doping, we observe clear signatures of Fermi pockets around these points. 

This is one of the few examples where  numerical simulations  show how quantum fluctuations can destroy a large-$N$ phase. 
\end{abstract}

\maketitle
\section{Introduction}

Spin nematic phases  are characterized by an absence of local magnetic moment and 
long range orientational ordering  of the spin degrees of freedom.  Such phases were first introduced by 
Andreev and Grishchuk~\cite{Andreev84} and then discovered in several magnetic models~\cite{Papanicolaou88,Chandra90}. 
Recently, spin nematic phases have also   
attracted attention in the the context of cold atoms  where a gas 
bose condensate of $^{52}Cr$ atoms in magnetic trap has been realized~\cite{Griesmaier05}. 
Due to the high spin, $J=3$, of the $^{52}Cr$ atoms, it has been suggested that,  
upon release of the spin degrees of freedom,  spin nematic states can be realized~\cite{Pfau06,Diener06}.
Mean-field realization of spin-nematic phases in particle-hole symmetric 
fermionic models on a square lattice  have equally been proposed by 
Nersesyan {\it et al.}~\cite{Nersesyan91}. Those phases are characterized by a checkerboard 
pattern of {\it alternating spin currents} around elementary plaquettes, and are coined 
spin flux phases (SFP).  They break $SU(2)$ spin symmetry as well as  the  lattice symmetry, 
but conserve time reversal symmetry. Such a phase could also possibly occur in an extended Hubbard model~\cite{Schulz89}. The goal of this 
paper is to provide a \emph{microscopic} realization of SFP. 

In this article, we concentrate on fermionic models on a square lattice and investigate
the stability of the spin-flux phase.  To do so, we consider  the multi-flavored  model 
Hamiltonian:
\begin{eqnarray}
\label{model.eq}
	H & = & - t \sum_{\langle i,j\rangle, \alpha} \left( c^{\dagger}_{i,\alpha} c_{j,\alpha} + h.c. \right) \nonumber \\
        &- & \frac{g}{2N} \sum_{\langle i,j\rangle } 
        \left( \sum_{\alpha} i c^{\dagger}_{i,\alpha} \vec{\sigma} c_{j,\alpha} 
         -i c^{\dagger}_{j,\alpha} \vec{\sigma} c_{i,\alpha} \right)^2 .
\end{eqnarray}
Here, the sum runs over nearest neighbors on a two-dimensional square lattice. The spinors 
$ c^{\dagger}_{i,\alpha} = \left( c^{\dagger}_{i,\uparrow,\alpha},c^{\dagger}_{i,\downarrow,\alpha}  \right)$
where $\alpha$ ranges from  $1$ to $N$. 
In the limit $N \rightarrow \infty$, the saddle point approximation becomes exact,  
and we recover the mean-field  results of Ref.~\onlinecite{Nersesyan91}.   
As $N$ is reduced, quantum 
fluctuations around this saddle point are progressively taken in account and ultimately at 
$N=1$ we recover a spin-1/2 fermionic Hamiltonian.  As we will see, the model, 
of Eq. (\ref{model.eq})  allows sign free auxiliary field quantum Monte Carlo (QMC) simulations for 
arbitrary band fillings and values of $	N$.  Hence, we can map out the phase diagram as a function of doping for various $N$, 
at fixed choice of coupling constant $g/t$. In the following, we fix $t=1$ as the energy unit.

The article is organized as follows.  In the next section, we briefly describe the 
path integral formulation of the partition function which is the basis of the mean field and 
quantum Monte Carlo simulations.  In section~\ref{Results} we present our numerical results  which
allow us to map out the phase diagram as a function of $N$ and doping for a  fixed coupling strength. 
Finally,  we summarize our results in section~\ref{Conclusions}.

\section{Mean-field and quantum Monte Carlo simulations. }

Both the saddle point equations and the QMC simulations  are based on a path 
integral formulation of the  partition function. Carrying out a Trotter breakup of the 
kinetic and 
interaction terms  as well as a Hubbard-Stratonovitch   transformation for the interaction term, 
the  partition function reads: 
\begin{widetext}
\begin{eqnarray}
    & &	Z \simeq  \int  
   \underbrace{\prod_{\langle i, j \rangle, \tau  } {\rm d} \vec{\Phi}_{\langle i, j \rangle}(\tau) }_{ \equiv D \vec{\Phi} } 
    e^{- \sum_{\langle i, j \rangle, \tau }   \vec{\Phi}_{\langle i, j \rangle}^2(\tau) /2 }
   \, {\rm Tr } \left[  \prod_{\tau} e^{ t \Delta \tau  \sum_{\langle i,j\rangle, \alpha} \left( c^{\dagger}_{i,\alpha} c_{j,\alpha} + h.c. \right) } \right. \nonumber \\ 
& &  \left. e^{\displaystyle \sqrt{ \Delta \tau g/N} \sum_{\langle i,j\rangle }  \vec{\Phi}_{\langle i, j \rangle}(\tau)\cdot
  \left( \sum_{\alpha} i c^{\dagger}_{i,\alpha} \vec{\sigma} c_{j,\alpha} 
         -i c^{\dagger}_{j,\alpha} \vec{\sigma} c_{i,\alpha} \right)
 } \right]   =  \nonumber \\
& & 
 \int D \vec{\Phi} e^{- \sum_{\langle i, j \rangle, \tau }   \vec{\Phi}_{\langle i, j \rangle}^2(\tau) /2 }
   \, {\rm Tr } \left[  \prod_{\tau} e^{ t \Delta \tau  \sum_{\langle i,j\rangle} \left( c^{\dagger}_{i} c_{j} + h.c. \right) }  e^{ \sqrt{ \Delta \tau g/N} \sum_{\langle i,j\rangle }  \vec{\Phi}_{\langle i, j \rangle}(\tau)\cdot
  \left(  i c^{\dagger}_{i} \vec{\sigma} c_{j} 
         -i c^{\dagger}_{j} \vec{\sigma} c_{i} \right)   } \right]^{N}
\end{eqnarray}
\end{widetext}
where 
$ c^{\dagger}_{i}   = \left( c^{\dagger}_{i,\uparrow }, c^{\dagger}_{i,\downarrow} \right) $. 
Note that in the last equation, the trace runs over  a single flavor.

As is well known, in the large $N$ limit the  mean-field solution is recovered. 
With the substitution
\begin{equation}
	\vec{\eta}_{\langle i, j \rangle}(\tau)  = \frac{1}{\sqrt{N \Delta \tau g} }  
      \vec{\Phi}_{\langle i, j \rangle}(\tau)
\end{equation}
one obtains
\begin{widetext}
\begin{eqnarray}
\label{HS.eq}
	& & Z \simeq 
  \int D \vec{\eta} e^{- N S(\eta) }  \\
	& & S(\eta) = 
   \sum_{\langle i, j \rangle, \tau }  \Delta \tau g 
        \vec{\eta}^2_{\langle i, j \rangle}(\tau) /2 - \ln 
   {\rm Tr }  \prod_{\tau} e^{ t \Delta \tau  \sum_{\langle i,j\rangle} \left( c^{\dagger}_{i} c_{j} + h.c. \right) }  
e^{  \Delta \tau g \sum_{\langle i,j\rangle }  \vec{\eta}_{\langle i, j \rangle}(\tau)\cdot
  \left(  i c^{\dagger}_{i} \vec{\sigma} c_{j} 
         -i c^{\dagger}_{j} \vec{\sigma} c_{i} \right)   }  \nonumber
\end{eqnarray}
\end{widetext}

In this large $N$ limit, the integral over the the fields $\eta$ is dominated by the saddle point configuration and  fluctuations around the saddle point are negligible. Moreover, 
by neglecting the $\tau$ dependence of the Hubbard-Stratonovitch fields, one obtains~:
\begin{eqnarray}
	& & S(\eta) = \beta g \sum_{\langle i,j\rangle} 
       \vec{\eta}_{\langle i, j \rangle}^2 /2 - \ln 
   {\rm Tr } e^{-\beta H_{MF}(\eta) } \nonumber \\
	& & H_{MF}(\eta) = - t \sum_{\langle i,j\rangle} \left( c^{\dagger}_{i} c_{j} + h.c. \right)\nonumber \\
  & &	- g   \sum_{\langle i,j\rangle }  \vec{\eta}_{\langle i, j \rangle}\cdot 
  \left(  i c^{\dagger}_{i} \vec{\sigma} c_{j} 
         -i c^{\dagger}_{j} \vec{\sigma} c_{i} \right)   
\end{eqnarray}
and the saddle-point equations read~:
\begin{equation}
	\frac{\partial S(\eta)  } { \partial \vec{\eta}_{\langle i,j \rangle}} = 0
\end{equation}
which correspond precisely to the mean-field solution~\cite{Nersesyan91}
\begin{equation}
	\vec{\eta}_{\langle i,j \rangle} = \langle  i c^{\dagger}_{i} \, \vec{\sigma} \, c_{j} 
         -i c^{\dagger}_{j} \, \vec{\sigma} \, c_{i}  \rangle_{H_{MF}}
\end{equation}

\subsection{Staggered spin flux mean field solution}
We consider the restricted order parameter: 
\begin{equation}
	\vec{\eta}_{\langle i,j \rangle} = \eta (-1)^{i_x + i_y} 
 ( \delta_{j-i,a_x} - \delta_{j-i,a_y} )\, \vec{e}_z.
\end{equation}
This choice is certainly valid at half-band filling where  perfect nesting pins the 
dominant instabilities to the wave vector $\vec{Q} = (\pi,\pi) $. Away from half-filling, 
order parameters at incommensurate wave vectors may be favorable. This is not taken into
account in the present Ansatz. 
With the above form, the mean-field Hamiltonian reads: 
\begin{widetext}
\begin{equation}
	H_{MF} = \sum_{\vec{k} \in MBZ, \sigma} 
\left( c^{\dagger}_{\vec{k},\sigma}, c^{\dagger}_{\vec{k} + \vec{Q},\sigma}  \right) 
\left( 
\begin{array}{cc}
	\varepsilon(\vec{k}) - \mu              &   -\sigma g \eta \overline{z(\vec{k})}  \\
      -\sigma g \eta z(\vec{k}) &    \varepsilon(\vec{k} + \vec{Q}) - \mu 
\end{array}
\right)
\left(
\begin{array}{c}
c_{\vec{k},\sigma} \\ c_{\vec{k} + \vec{Q},\sigma} 
\end{array}
\right)
\end{equation}
\end{widetext}
where the sum over $\vec{k}$ is restricted to the magnetic Brillouin zone (MBZ) and 
with 
\begin{equation}
	z({\vec{k}}) = -2i \left( \sin ( k_x + Q_x/2 ) - 
           \sin ( k_y + Q_y/2 )  \right) 
\end{equation}
Thus, we obtain the saddle point equation
\begin{eqnarray}
	\eta = & & \frac{1}{2 L^2} \sum_{\vec{k} \in MBZ, \sigma}  
\langle 
 \left( c^{\dagger}_{\vec{k},\sigma}, c^{\dagger}_{\vec{k} + \vec{Q},\sigma}  \right)  
\nonumber \\
& & \left( 
\begin{array}{cc}
	       0                    &   \sigma  \overline{z(\vec{k})}  \\
      \sigma  z(\vec{k}) &       0 
\end{array}
\right)
\left(
\begin{array}{c}
c_{\vec{k},\sigma} \\ c_{\vec{k} + \vec{Q},\sigma} 
\end{array}
\right)
\rangle
\end{eqnarray}
which we solve self-consistently. 

\begin{figure}[!ht]
\includegraphics[width=7cm]{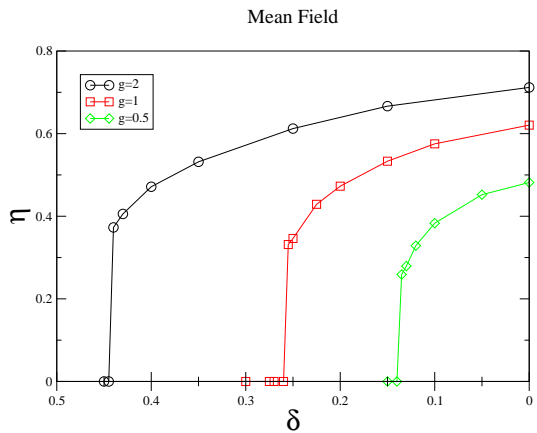} 
\caption{(Color online) Mean Field order parameter as  a function of band-filling for various coupling strengths $g$. }
\label{MF.fig}
\end{figure}

\begin{figure}[!ht]
\includegraphics[width=8cm]{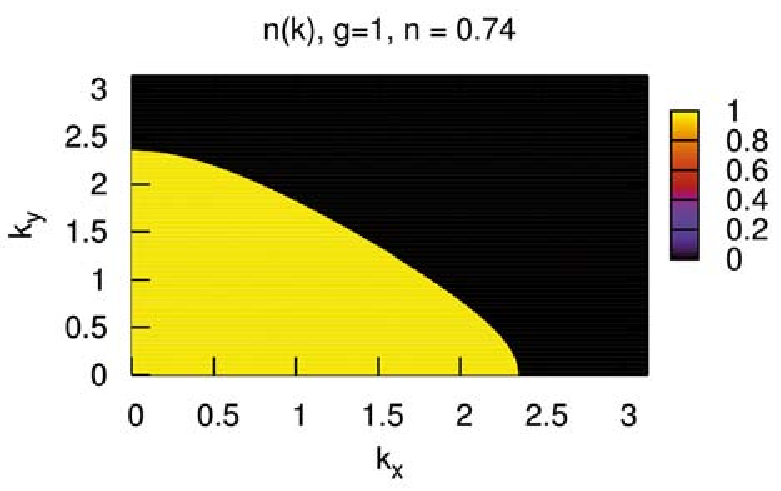} 

\includegraphics[width=8cm]{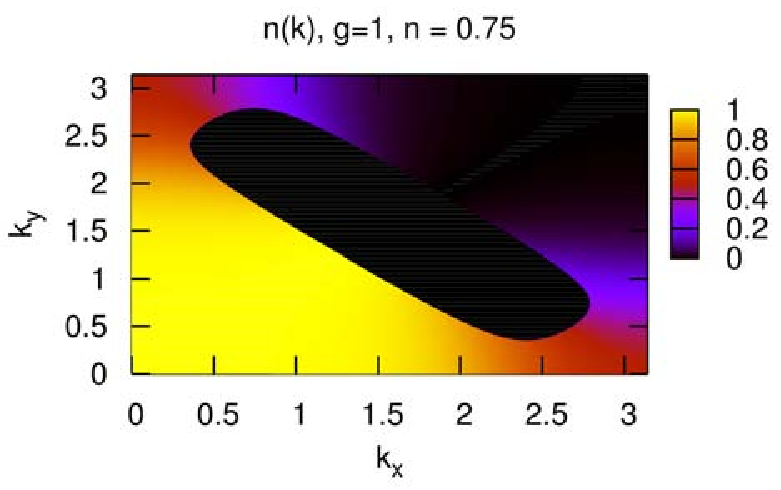} 

\includegraphics[width=8cm]{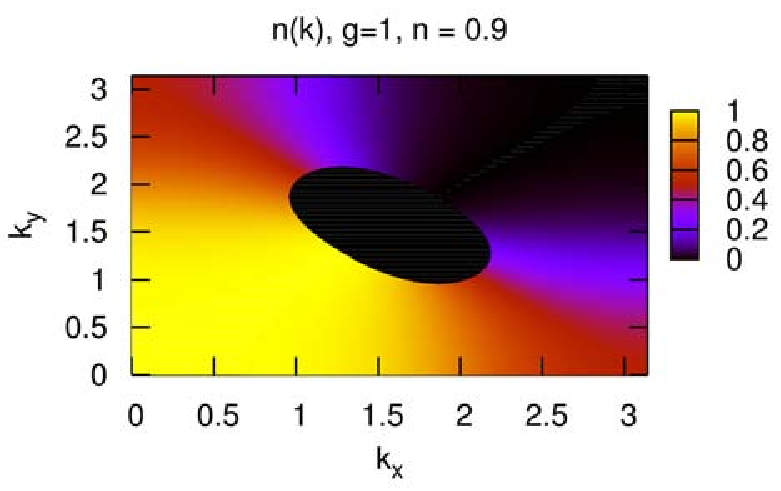}

\includegraphics[width=8cm]{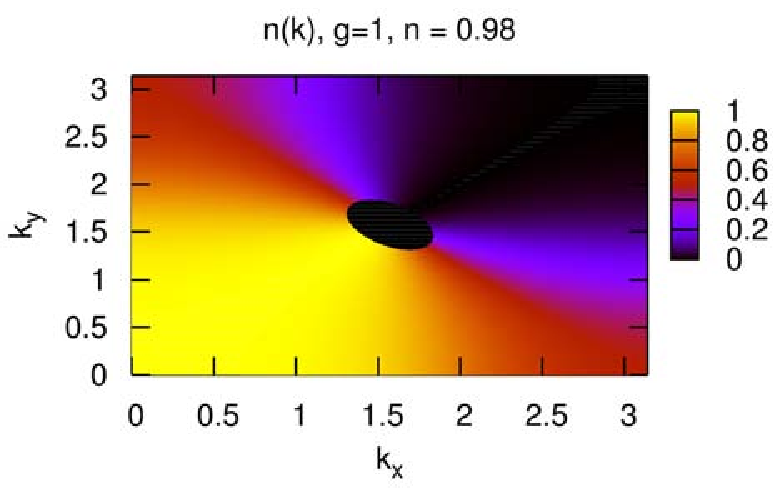}
\caption{(Color online) Mean-field solution of $n(\vec{k})$ over the Brillouin zone as a function of band filling for a fixed $g=1$.}
\label{NkMF.fig}
\end{figure}

Figure~\ref{MF.fig} shows the order parameter as a function of doping. The staggered spin flux 
phase survives up to a finite critical doping, where a first order transition to the paramagnetic 
phase occurs. This transition is signaled by a jump 
in the order parameter.    At half-band filling, the   single
particle dispersion relation is given by, 
\begin{equation}
	E(\vec{k}) = \pm \sqrt{  \varepsilon^2(\vec{k})  + \Delta^2(\vec{k}) }   
\end{equation}
with $\Delta(\vec{k}) = 2 g \eta \left( \cos(k_x)  - \cos(k_y) \right)$. Hence, it exhibits 
Dirac cones around the $( \pm \pi/2, \pm \pi/2)$   $k$-points and  the Fermi {\it  surface} 
is given by those four points. In the doping range where the order parameter 
does not vanish  the Fermi surface consists of hole-pockets centered around the 
above mentioned $\vec{k}$-points (See Fig.~\ref{NkMF.fig}).  Note that 
due to the d-wave symmetry of the order parameter, the Fermi surface is not invariant 
under reflections across the $(1,1)$ axis 
(i.e. $k_x \rightarrow k_y $ and $k_y \rightarrow k_x $). 
As appropriate for a first order transition,  the Fermi surface changes abruptly from hole 
pockets around  $(\pi/2,\pi/2)$ in the spin flux phase to a 
large  Fermi surface centered around $ (0,0) $ in the paramagnetic phase. 

\subsection{Quantum Monte-Carlo}

The Hamiltonian occurring in~(\ref{HS.eq}) is quadratic in the fermionic variables so that 
the trace can be carried out analytically to obtain:
\begin{equation}
Z  \simeq \int  D \vec{\Phi}   
e^{- \sum_{\langle i, j \rangle, \tau }   \vec{\Phi}_{\langle i, j \rangle}^2(\tau) /2 }
\left[ {\rm det}  M(\Phi)\right]^N
\end{equation}
Since the spin current is even under time reversal symmetry, it has been shown that,  
when $g\geq 0$,  the 
fermionic determinant is positive for each Hubbard-Stratonovitch configuration~\cite{Wu04}.    
This absence of sign problem is valid for any lattice topology and any filling. 
Hence  each configuration of Hubbard Stratonovitch fields 
can be sampled  according to its weight with Monte Carlo techniques.

In the following, we study the ground-state properties of the Hamiltonian~(\ref{model.eq}) with 
the projector
auxiliary field QMC algorithm on a two-dimensional square lattice. The details on this algorithm can be found for example in 
Ref.~\onlinecite{Capponi00}, 
where the authors consider a very similar model from the technical point of view. Dynamical 
information is obtained by using  a recent implementation of the  stochastic analytical 
continuation~\cite{Beach04a,Sandvik98}.

\section{Numerical results}\label{Results}
\subsection{Phase diagram}
For $N=1$,  the interaction term of our model can be rewritten, up to a constant as: 
\begin{eqnarray}
\label{Interaction_N1}
	&-& \frac{g}{2} \sum_{\langle i,j\rangle } 
        \left( \sum_{\alpha} i c^{\dagger}_{i,\alpha} \vec{\sigma} c_{j,\alpha} 
         -i c^{\dagger}_{j,\alpha} \vec{\sigma} c_{i,\alpha} \right)^2  \nonumber \\
         =  
         &-&3g\sum_{\langle i,j\rangle } (c^{\dagger}_{i,\uparrow}c^{\dagger}_{i,\downarrow}c^{\phantom{\dagger}}_{j,\downarrow} c^{\phantom{\dagger}}_{j,\uparrow}
  + h.c.)  \nonumber \\
&+&\frac{3}{2}g \sum_{\langle i,j\rangle } (n_i-1)(n_j-1) \nonumber \\
&-&2g\sum_{\langle i,j\rangle } \vec{S}_i \cdot \vec{S}_j
\end{eqnarray}
As apparent, there is a pair-hopping term so that we can expect an on-site superconducting
 instability which we pick up by measuring 
the equal time pair correlation functions: 
\begin{equation}
SC(\vec{R})=\langle c^\dagger_{\vec{R}\uparrow} c^\dagger_{\vec{R}\downarrow} c^{\phantom{\dagger}}_{0\downarrow} c^{\phantom{\dagger}}_{0\uparrow} \rangle.
\end{equation}
In the presence of long range off-diagonal order,  its Fourier transform at $\vec{Q}=(0,0)$
\begin{equation}\label{SC_Q}
SC(\vec{Q}=(0,0))=\frac{1}{L^2}\sum_{\vec{R}} SC(\vec{R})
\end{equation}
should converge to a finite value in the thermodynamic limit, whereas it vanishes as $1/L^2$ in the absence of long-range order.
\begin{figure}[h]
\includegraphics[width=8cm,clip]{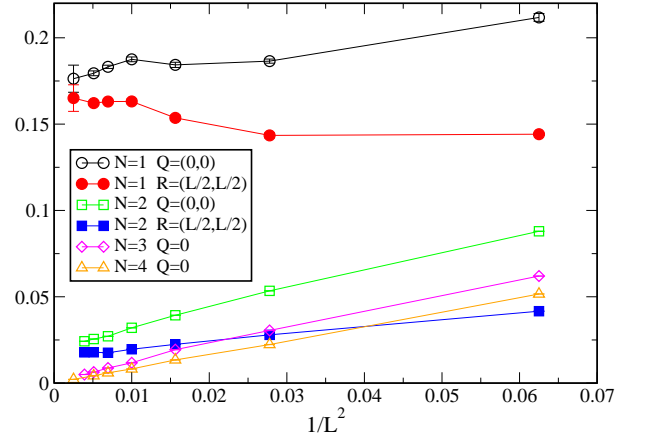} 
\caption{(Color online) Finite-size scaling of the superconducting correlations vs the inverse number of sites $1/L^2$ of the system 
($L$ is the linear size) for the 
half-filled case and different $N$. The $\vec{Q}=(0,0)$ Fourier correlations are plotted (open symbols) and, when long range order is present, 
we also show the largest distance real-space correlations (filled symbols).} 
\label{supra_d0.fig}
\end{figure}
 
The order parameter of the spin flux phase expected  in the limit of $N \rightarrow \infty $, 
reads: 
\begin{eqnarray}\label{sfp.eq}
\langle  \left[ \vec{J}^s(\vec{R},\vec{R}+\vec{x})-\vec{J}^s(\vec{R},\vec{R}+\vec{y}) \right]  
\nonumber \\
   \cdot \left[ \vec{J}^s(\vec{0},\vec{0}+\vec{x})-\vec{J}^s(\vec{0},\vec{0}+\vec{y}) \right]
       \rangle
\end{eqnarray}
where 
\begin{equation}
\vec{J}^s(\vec{R},\vec{R}+\vec{x})= i \left[ 
   c^\dagger_{\vec{R},\mu} \vec{\sigma}_{\mu \nu} c^{\phantom{\dagger}}_{\vec{R}+\vec{x},\nu} 
  -    
  c^\dagger_{\vec{R}+\vec{x},\mu} \vec{\sigma}_{\mu \nu} c^{\phantom{\dagger}}_{\vec{R},\nu} 
   \right]
\end{equation}
is the bond spin current operator along $x$ (summation over repeated indices is assumed). A similar definition holds along the $y$ direction. 
As expected from the mean-field solution, we have observed that the strongest signal occurs in 
the d-wave symmetry channel and the Fourier transform of these  correlations is maximum at 
$\vec{Q}=(\pi,\pi)$. 
\begin{figure}
\includegraphics[width=8cm]{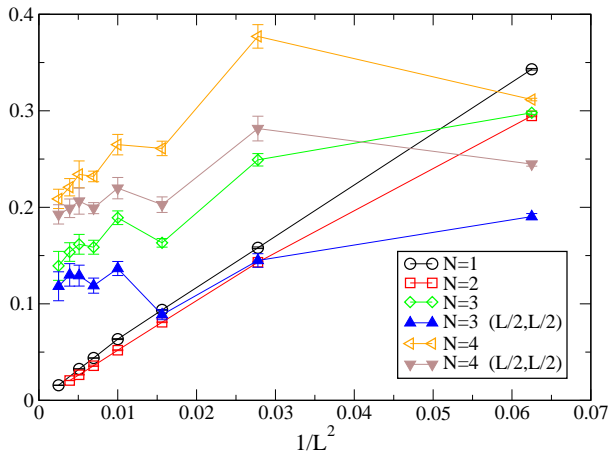} 
\caption{(Color online) Similar as the previous plot for $d$-wave spin current correlations.}
\label{jspind_d0.fig}
\end{figure}

\begin{figure}
\includegraphics[width=0.4\textwidth,]{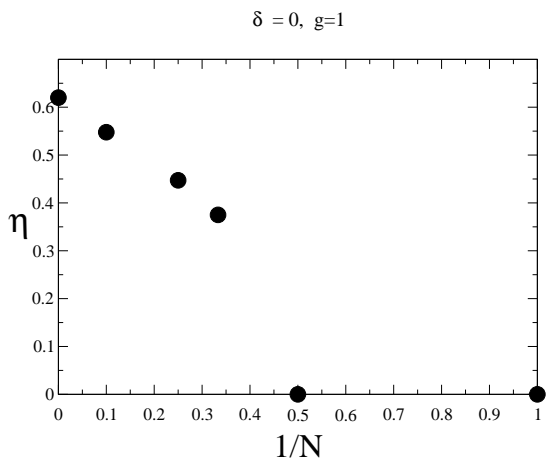} 
\caption{Order parameter $\eta$ as a function of $N$ at half-filling. The data point at $N= \infty$ corresponds to
the mean-field value. }
\label{eta_N}
\end{figure}

\begin{figure}
\includegraphics[width=8cm]{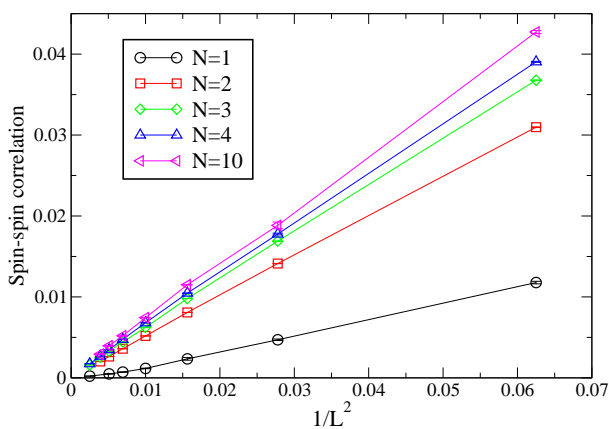} 
\caption{(Color online) Finite-size scaling of the spin-spin correlations vs the inverse 
volume,  $1/L^2$, of the system for the 
half-filled case and different $N$. For all cases, the $\vec{Q}=(\pi,\pi)$ Fourier correlations (which is the largest one) vanish in the thermodynamic limit, indicating the absence
of magnetic order.}
\label{sz.fig}
\end{figure}

{\it Technical details:} we are
able to simulate up to $20\times 20$ square lattices with 
a projection parameter $\theta t =10$. We fix $\Delta \tau=0.1$ for the
Trotter decomposition. For simplicity, in the following, we fix the coupling constant $g=1$. 

{\it Half-filling case~:}
On Fig.~\ref{supra_d0.fig}, we plot the scaling of the superconducting correlations for different values of $N$. According to our definition~(\ref{SC_Q}), 
the $\vec{Q}=(0,0)$ 
correlations are divided by the number of sites so that a finite value in the thermodynamic limit indicates long range order. 
Moreover, when this is the case, we also plot the largest distance real-space correlations that should converge to the same
value in the thermodynamic limit. In particular, we have a clear signal of 
s-wave superconducting phase for $N=1$ and $N=2$ but this phase disappears for larger $N$. 

At $N=1$ and under the canonical transformation,
\begin{equation}
      c^{\dagger}_{i,\downarrow}   \rightarrow  (-1)^i c_{i,\downarrow} \; \;  \; \; 
      c^{\dagger}_{i,\uparrow }    \rightarrow  (-1)^i c^{\dagger}_{i,\uparrow}  
\end{equation}
the interaction term transforms as, 
\begin{eqnarray}
-  \frac{g}{2}  
        \left( i c^{\dagger}_{i} \vec{\sigma} c_{j} 
         -i c^{\dagger}_{j} \vec{\sigma} c_{i} \right)^2  \rightarrow   
        6g \vec{S}_i \cdot \vec{S}_j   & &  \nonumber \\
	- \frac{g}{2} (n_i -1) (n_j -1) +g ( c^{\dagger}_{i,\uparrow} c^{\dagger}_{i,\downarrow} c^{\phantom{\dagger}}_{j,\downarrow} 
c^{\phantom{\dagger}}_{j,\uparrow} + h.c.),
\end{eqnarray}
and the kinetic energy remains invariant.  The observed long-ranged  
pairing correlations  map onto  long ranged tranverse spin-spin correlations. Since the 
transformed model has an $SU(2)$ spin symmetry,  the transverse spin-spin correlations  on 
any finite lattice take the same value as the longitudinal spin-spin correlations. 
Transforming back to the original model maps the longitudinal spin-spin correlations to 
a charge density wave order. Hence at half-filling and $N=1$ pairing correlations and 
charge density wave correlations are locked in by symmetry.   
Clearly, doping  breaks this symmetry~\footnote{Precisely the same transformation has 
been used to 
investigate the half-filled attractive Hubbard model~\cite{Scalettar89}.}.

\begin{figure}
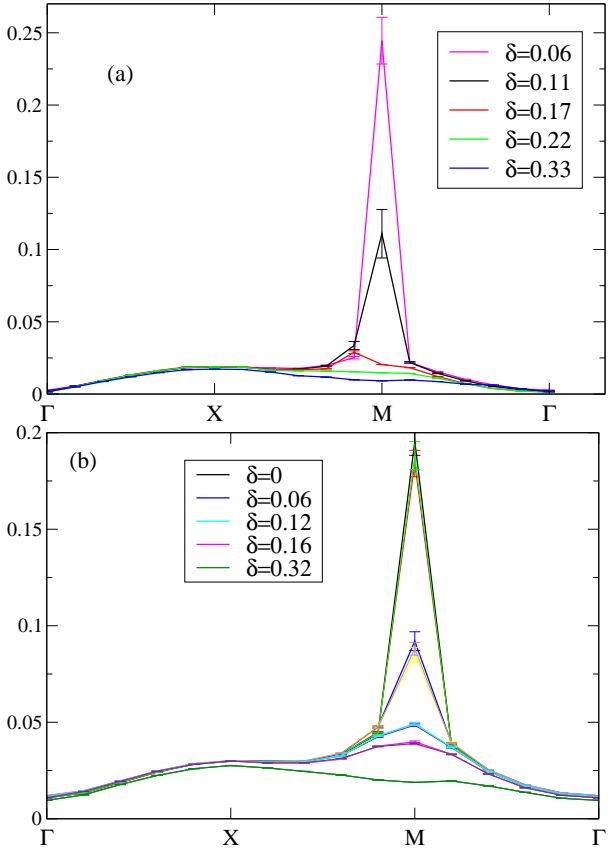

\includegraphics[width=8cm]{jspind_N10L12} 
\includegraphics[width=8cm]{jspind2_N3L10}
\caption{(Color online) (a) Spin current correlations, averaged over the 3 components, 
vs ${\vec{k}}$ along the $(\Gamma X M)$ path in the Brillouin zone. 
Parameters are $L=12$, $N=10$ and $g=1$. As a function of doping $\delta$ (increasing from top to bottom), the maximum at ($\pi,\pi$) is rapidly suppressed. 
(b) Same for $L=10$ and $N=3$. Here, we plot all 3 components showing that $SU(2)$ symmetry is recovered in our data. 
As a function of doping (increasing from top to bottom), 
the maximum at ($\pi,\pi$) is rapidly suppressed.}
\label{jspind_doping.fig}
\end{figure}

Similarly, we investigate the occurrence of SFP phase by computing the
corresponding correlations (see Eq.~(\ref{sfp.eq})). Moreover, for a finite system, the $SU(2)$
symmetry of these correlations cannot be broken so that we average
over the 3 components to reduce the error bars. For example, we plot on Fig.~\ref{jspind_doping.fig}~(b) all three components of the
SFP correlations showing that indeed $SU(2)$ symmetry is restored (within the error bars) despite being 
explicitely broken for a given Hubbard-Stratonovitch configuration.
 On Fig.~\ref{jspind_d0.fig}, we plot the scaling of these correlations
showing that SFP is present in the thermodynamic limit  for $N\ge 3$ but is not stable for smaller
values of $N$. Again, when the $\vec{Q}=(\pi,\pi)$ Fourier component
indicates long range order (LRO), we also plot the scaling of the largest distance
correlations that should have the same value in the thermodynamic
limit for an ordered phase.  In particular, we recover the existence
of SFP phase for large $N$ as found at the mean-field level. A non
trivial result is that this phase survives for \emph{finite} $N\ge 3$ since
our QMC simulations, which are free of the sign problem, include
quantum fluctuations. 
We observe rather strong finite size effects and in particular, the signal is weaker when 
the lattice contains
$(\pi/2,\pi/2)$ k-points in its Brillouin zone. Indeed, as we will discuss in the 
following, this point correspond to the low-energy excitations.  
Since the   staggered spin current correlation function   of Fig.~\ref{jspind_d0.fig} 
converges to the square of the 
order parameter $\eta$, we can extract this quantity as a function of $N$.   Our results 
are plotted in Fig.~\ref{eta_N}. As apparent, the Monte Carlo results at finite values of $N$ 
smoothly scale to the mean field result valid at $N = \infty$.

In the SFP at $N \geq 3$, we have computed the spin-spin correlation functions 
$\langle \vec{S}_{ \vec{i}} \cdot \vec{S}_{ \vec{j}}  \rangle$.  
As shown on Fig.~\ref{sz.fig}, our results show no sign of long-range spin ordering and hence the absence of a magnetic moment. 
This  confirms the point of 
view that the spin-flux phase that we observe for $ N \geq 3 $ indeed corresponds to a 
\emph{spin nematic phase}. 

\begin{figure}
\includegraphics[width=8cm]{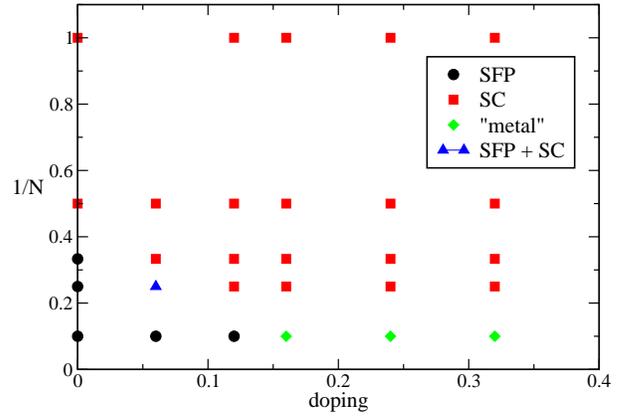} 
\caption{(Color online) Phase diagram as a function of doping and $1/N$ obtained with QMC simulations. At large
$N$, we recover the existence of a spin flux phase (SFP) over a finite doping region. For larger doping or 
smaller $N$, the system develops s-wave superconducting correlations (SC) or stays in a metallic phase (see text).}
\label{phasediag.fig}
\end{figure}

{\it Doping~:}
Since the SFP instability occurs in the particle-hole channel, it is favored at half-filling 
where the Fermi surface exhibits perfect nesting. But the question
about its existence under doping remains open. 
Our QMC simulation being free of the sign-problem for any filling, 
we have computed the phase diagram as a function of doping for different values of $N$. 
As expected from the mean-field results, we show on Fig.~\ref{phasediag.fig} that the SFP 
phase has a finite extension up to a critical doping.  In particular, when $N=10$, we find 
a critical doping around 0.15. 

On Fig.~\ref{jspind_doping.fig}, we plot the spin current correlations vs ${\vec{k}}$ along the $(\Gamma X M)$ path in the Brillouin zone (shown in  
Fig.~\ref{brillouin.fig}). 
For doping $\delta$ smaller than a critical value, we see a large maximum at ($\pi,\pi$), whereas for larger doping, the correlations become featureless,
indicating the disappearance of SFP order. We also observe that the maximum of the spin-current correlations move away from $(\pi,\pi)$:
for example, when $L=12$, $N=10$ and with doping $\sim 0.17$, the maximum is at $(5\pi/6,\pi)$ and equivalent k-points. 

By performing a finite-size scaling analysis of our data, we are able to draw conclusion 
about the existence or not of a SFP in the thermodynamic limit
as a function of doping and $N$. These results are summarized in the phase diagram 
shown in Fig.~\ref{phasediag.fig}. Several comments are in order~:  \\
{\bf i)} Under doping the SFP  only survives for large values of $N$, hence showing that it 
is very sensitive to fluctuations.   At $N=10$ the 
critical doping, $\delta_c \simeq 0.15 $  at which the SFP  vanishes is 
substantially  smaller than the $N \rightarrow \infty $ restricted mean-field result 
$\delta_c \simeq 0.25$.  In the vicinity of the phase transition from the  SFP to the 
paramagnetic phase transition at  $N \rightarrow \infty$, fluctuations will play an important 
role. Hence at large but finite values of $N$, substantial differences can be expected. 
Furthermore, our restricted mean-field solution  does not allow for  incommensurate 
ordering as suggested by the Monte Carlo results. \\
{\bf ii)} The rest of the phase diagram is  dominated by an s-wave  SC phase. At 
$N=1$,  the s-wave SC persists of course away from half-filling since it 
is a particle-particle instability that does not require any nesting property.
As $N$ grows, the s-wave SC order parameter is reduced and ultimately vanishes  to 
produce the paramagnetic phase in the $N \rightarrow \infty $ limit. 
At $N=10$  and $\delta > \delta_c \simeq 0.15$  we observe no SFP correlations and no 
SC correlations either.  Therefore, it seems that we might  have a metallic phase with 
a Fermi surface.  However, due to the presence of pair hopping in the model, we know that
this metallic phase will ultimately be unstable at very low temperature toward s-wave 
superconductivity. This instability probably occurs at too low temperature to be observed 
in our simulations.  To confirm this point of view, let us write the interaction term as: 
\begin{eqnarray}
	-\frac{g}{2N} \sum_{b} \left( \sum_{\alpha=1}^{N} \vec{J}^{s}_{b,\alpha}\right)^2 = 
	-\frac{g}{2} \sum_{b} \sum_{\beta=1}^{N} \nonumber \\
        \left[ \left(  \frac{1}{N}\sum_{\alpha\neq \beta}  \vec{J}^{s}_{b,\alpha} \right)  
        \vec{J}^{s}_{b,\beta }  + 
        \frac{1}{N} \vec{J}^{s}_{b,\beta} \cdot \vec{J}^{s}_{b,\beta}\right]
\end{eqnarray}
where $b=\langle \vec{i},\vec{j}\rangle $ denotes a nearest neighbor bond, and 
$ \vec{J}^{s}_{b,\alpha} $ is the spin current on bond $b$ for flavor index $\alpha$. 
This forms shows that in the large-N limit, the flavor index $\beta$  is embeded in the 
{\it mean field } spin current of  the other  flavors,  
$ \frac{1}{N}\sum_{\alpha\neq \beta}  \vec{J}^{s}_{b,\alpha} $. 
The second term, $ \frac{1}{N} \vec{J}^{s}_{b,\beta} \cdot \vec{J}^{s}_{b,\beta}$ is a 
factor $1/N$ smaller than the {\it mean-field } term and provides the pair hopping term as
explicitly shown in Eq.~(\ref{Interaction_N1}).  At $N = \infty $, only the mean-field 
term survives. For large but finite values of $N$,  a small pairing term is present and 
will in the paramagnetic phase trigger a superconducting instability at low temperatures. \\
{\bf iii)}  The coexistence a spin flux phase and superconductivity at finite doping 
cannot be excluded.  As we will see below, doping the SFP  leads to a Fermi 
surface consisting of hole pockets centered around the 
$(\pm \pi/2,\pm \pi/2)$ points in the Brillouin zone. Due to the presence of pair 
hopping, this Fermi surface should be unstable towards a superconducting state. As a result, 
for intermediate values of $N$ such that we observe SFP at zero doping and such that the superconducting 
signal at finite doping is not strongly reduced due to a $1/N$ factor, we have found numerical evidence that supports coexistence of both phases
for small doping (for instance, $N=4$ and $\delta\sim 0.06$ in the phase diagram).

\subsection{Dynamical properties}

To study the charge degrees of freedom, we compute the single particle spectral function
$A(\vec{k},\omega)$ which is related to the imaginary time Green function via:
\begin{equation}
       \langle  c^{\dagger}_{\vec{k},\alpha}(\tau)     c_{\vec{k},\alpha} \rangle
     = \frac{1}{\pi}
    \int_{0}^{\infty} {\rm d} \omega  e^{-\tau \omega} A(\vec{k}, -\omega).
\label{AKOM}
\end{equation}

\begin{figure}
\includegraphics[width=4cm]{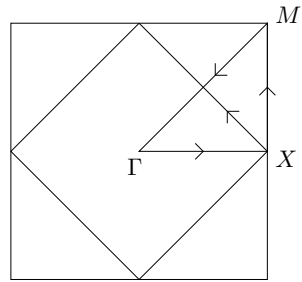} 
\caption{In the following spectral function, from bottom to top, the $k$ value follows the path $\Gamma$, $X$, $M$ and then, along the $(\pi,0)$ to $(0,\pi)$ line.}
\label{brillouin.fig}
\end{figure}

\begin{figure}
\includegraphics[width=8cm]{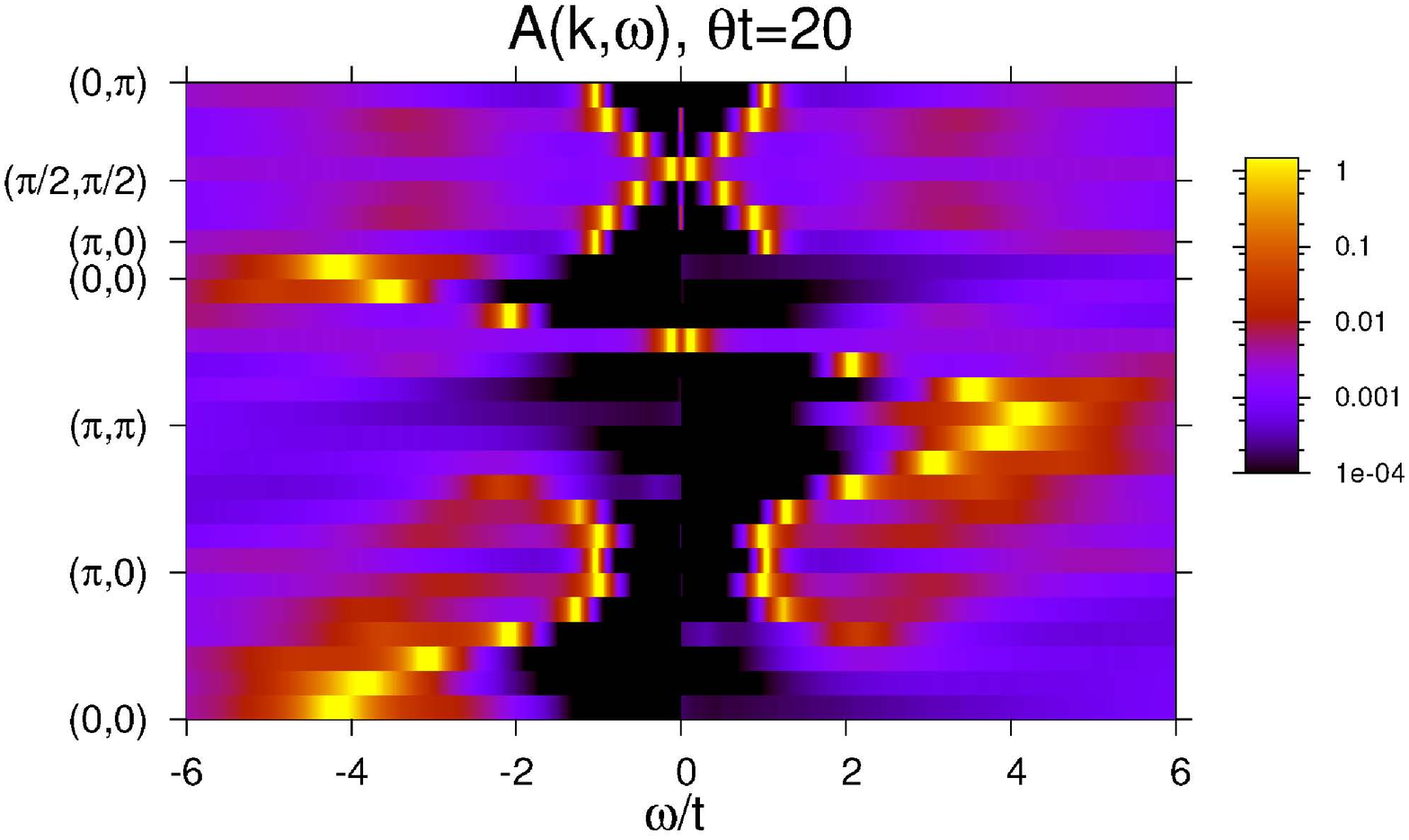} 

\includegraphics[width=8cm]{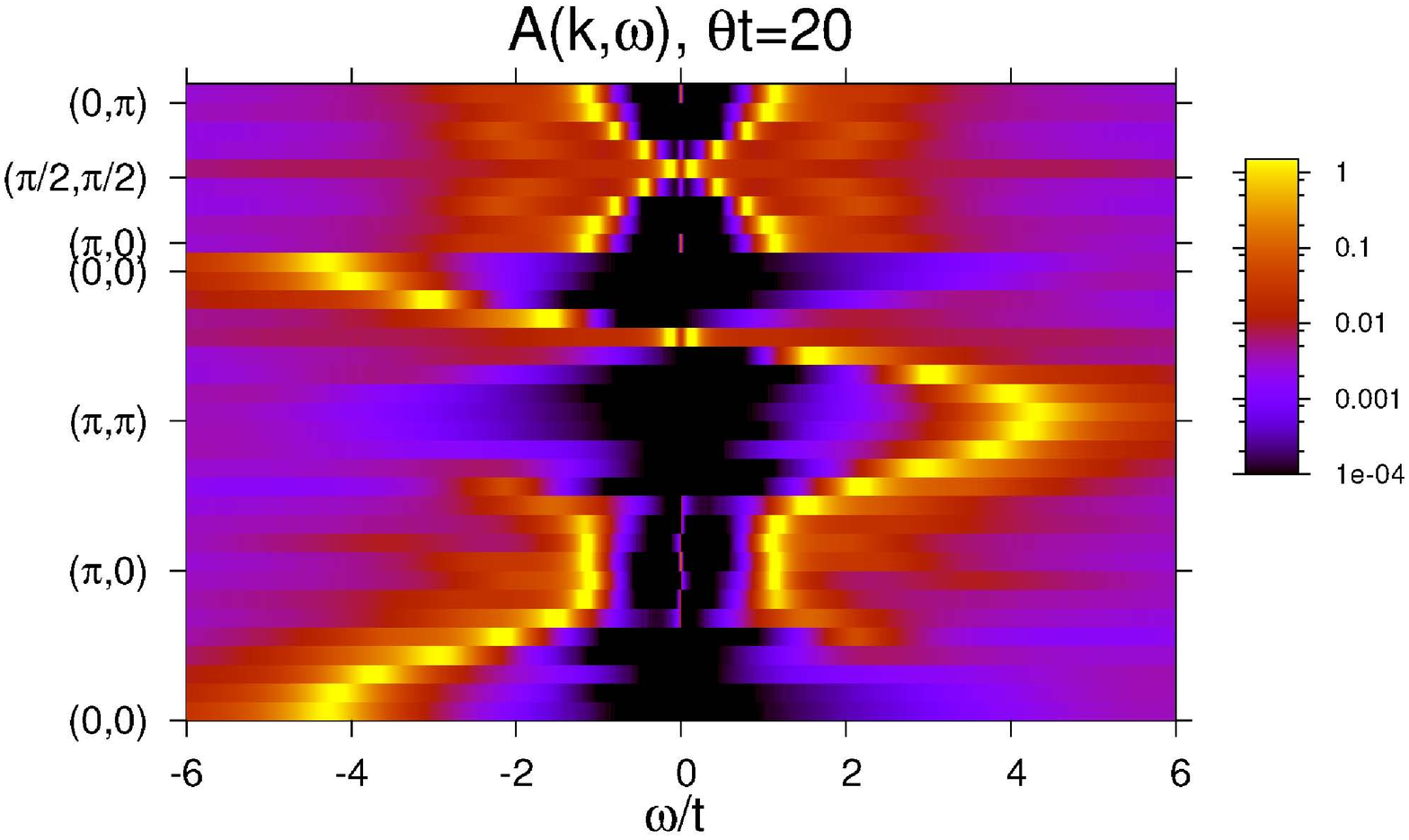} 

\caption{(Color online) Spectral functions vs frequency $\omega$ for different $\vec{k}$ points at half-filling. Up: $N=10$ and $L=12$; 
bottom: $N=4$ and $L=16$. Both set of parameters correspond to SFP and Dirac cones 
around $(\pi/2,\pi/2)$ are clearly seen. }
\label{Akw_SFP_d0.fig}

\end{figure}

In order to plot this single particle spectral function, we follow a path in the Brillouin 
zone shown on Fig.~\ref{brillouin.fig}.
In Fig.~\ref{Akw_SFP_d0.fig} we first plot two examples of 
$A(\vec{k},\omega)$ in the SFP at half-band filling. As expected from the mean-field 
calculation we observe a clear sign of the Dirac cones around the $ (\pi/2,\pi/2)$ 
${\vec{k}}$-points  in the Brillouin zone.  The overall dispersion relation  compares favorably
to the mean-field result 
\begin{equation}\label{meanfield_dispersion.eq}
	E(\vec{k}) = \pm \sqrt{  \varepsilon^2(\vec{k})  + \Delta^2(\vec{k}) }   
\end{equation}
with $\Delta(\vec{k}) = 2 g \eta \left( \cos(k_x)  - \cos(k_y) \right)$.

\begin{figure}
\includegraphics[width=8cm]{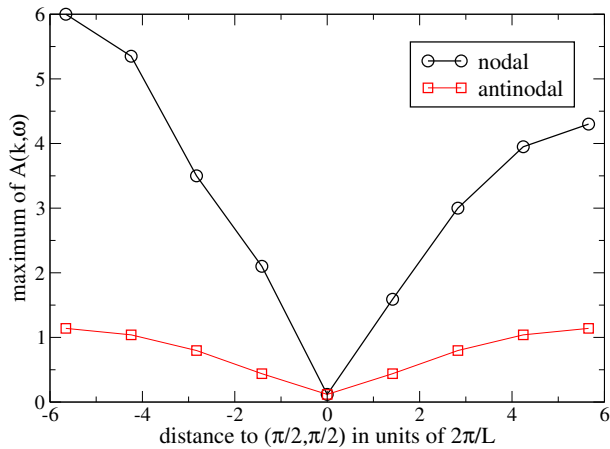} 

\caption{(Color online) Dispersion relation obtained with $N=4$ and $L=16$ at half-filling. The energy is taken at the maximum of the spectral function and data
are plotted vs distance to $(\pi/2,\pi/2)$. The small asymmetry is due to the uncertainty in the spectral functions obtained with 
a Maximum Entropy technique.  }
\label{dirac.fig}
\end{figure}

In order to be more quantitative, we plot the dispersion relation of the main branch (by looking at the maximum of
the spectral function) as a function of the distance to the nodal point. On Fig.~\ref{dirac.fig}, we clearly see the Dirac cone
structure with different velocities parallel and perpendicular to the Fermi surface. By fitting the slopes, we obtain for this set
of parameters, $v_\perp=3.11$ and $v_\parallel=0.578$. At the mean-field level, according to Eq.~(\ref{meanfield_dispersion.eq}), 
these two velocities are equal respectively to $2\sqrt{2}$ and $2\sqrt{2}g\eta$ so that their ratio gives directly access to the SFP order parameter $\eta$ (when 
$g=1$). Of course, the QMC data include some renormalization of these values and 
 the ratio of $v_\parallel / v_\perp =0.186$ is inconsistent with the SFP 
order parameter ( $ \eta \simeq 0.45$)  as extracted from the spin current 
correlations (see Fig.~\ref{jspind_d0.fig}).

\begin{figure}
\includegraphics[width=8cm]{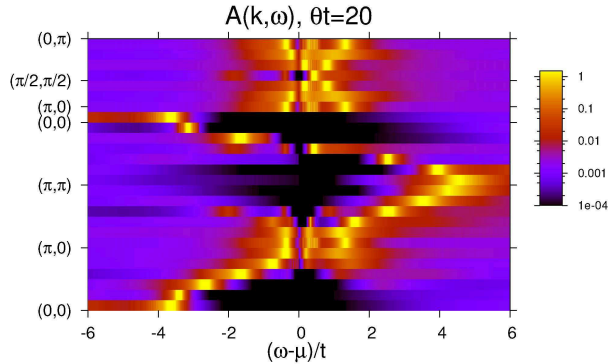} 
\caption{(Color online) Spectral functions vs $\omega-\mu$ for different $k$ points. Parameters are: $N=10$, $L=12$
and doping $0.12$. It corresponds to a doped SFP phase and Dirac cones around $(\pi/2,\pi/2)$ are clearly seen. The
chemical potential is $\mu=-0.48$.} 
\label{Akw_SFP_d0_12.fig}
\end{figure}

In the mean-field approach, doping the SFP leads to hole pockets centered around the 
$(\pi/2,\pi/2)$ $\vec{k}$-points. Numerical QMC simulations confirms this as shown on Fig.~\ref{Akw_SFP_d0_12.fig}. At $N=10$ and
$\delta = 0.12$  we observe a clear signature of the Dirac cone, however the quasiparticle 
at $\vec{k} = (\pi/2,\pi/2) $ lies above the Fermi energy.    

\begin{figure}
\includegraphics[width=8cm]{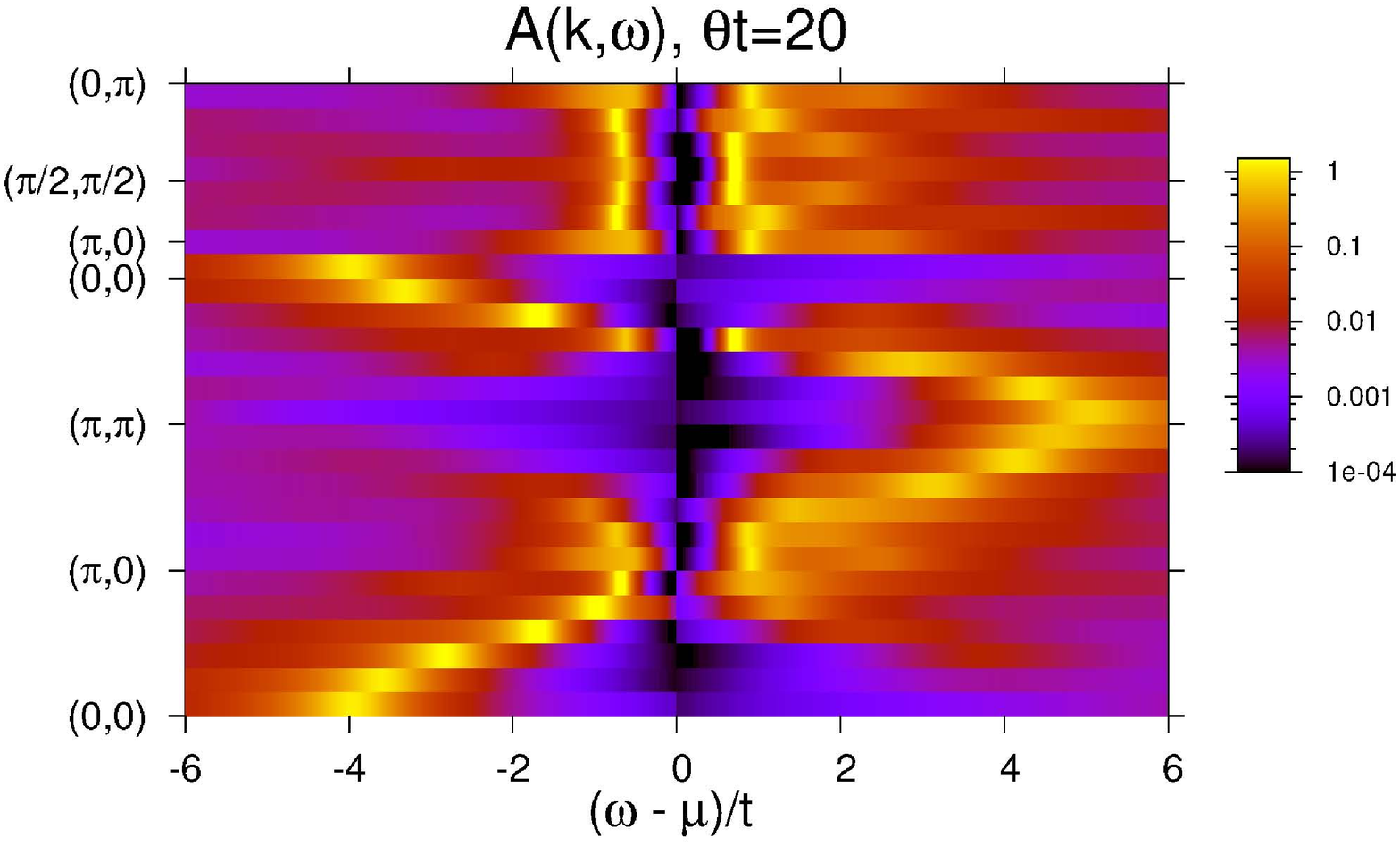}

\includegraphics[width=8cm]{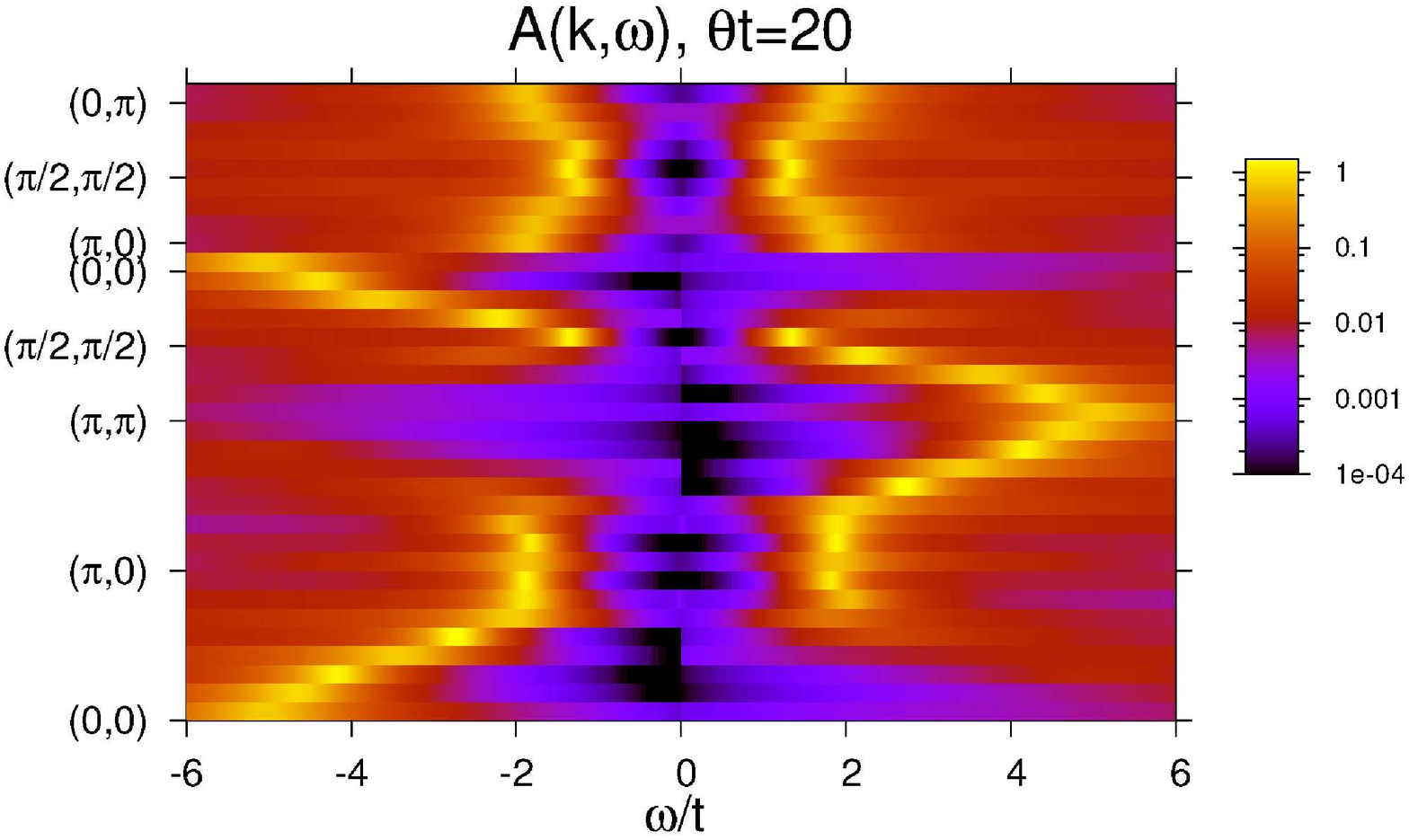}

\caption{(Color online) Spectral functions vs frequency $\omega$ for different $k$ points. 
Top: $N=3$ and $L=12$ for doping $0.12$ ($\mu=-0.7$).
Bottom $N=2$ and $L=16$ at half-filling ($\mu=0$).  
Both set of parameters correspond to s-wave superconducting phase. }
\label{Akw_SC.fig}
\end{figure}

Deep  in the superconducting phase,  the single particle dispersion relation follows the 
the mean-field form: 
\begin{equation} 
      E(\vec{k}) = \pm \sqrt{ ( \epsilon(k) - \mu)^2  + \Delta_{sc}^2} 
\end{equation}
with $k$-independent superconducting gap.  Such a dispersion relation is consistent with 
the Monte Carlo data of Fig. \ref{Akw_SC.fig} at $ N=3 $ and $\delta = 0.12$.  In the  
proximity  of the SFP at $N=2$ and half-band filling  (see Fig. \ref{Akw_SC.fig}) 
the dispersion  is not captured by the  above form. In particular, a  modulation of the 
gap along the $(\pi,0)$ to $(0,\pi)$ line is observed. This modulation is reminiscent of the 
Dirac cone structure observed in the spin-flux phase, and hence allows the  interpretation
that  short range spin currents survive in the superconducting state in the  proximity of the 
SFP. 

\section{Conclusion}
\label{Conclusions}
We have shown the existence of a spin nematic phase in a two-dimensional electronic model.  This
phase had been proposed at the mean-field level and it is the aim of this article to 
investigate its  stability against 
quantum fluctuations. In order to do so,  we have performed unbiased 
QMC simulations (without sign problem) for a variety 
of models with a flavor parameter $N$.  With this trick, we can interpolate from the large $N$ 
limit where the mean-field is valid to finite $N$ regime where quantum fluctuations  around the 
saddle point are progressively  taken into account. 
We find that the spin-flux phase is stable for a range of finite $N$ but is ultimately destroyed 
for the more realistic $N=1$ case. This is one of the few examples
where QMC simulations are able to show how quantum fluctuations can destroy a large-$N$ phase~\cite{Assaad04}. 
Since our model includes pair hopping processes, 
when the spin-flux phase is found to be unstable, it is replaced by an on-site $s$-wave superconductivity. 

We have also investigated the phase diagram as a function of doping since our QMC simulations 
are free of the sign problem for any filling. Because the spin-flux
phase is due to the nesting property of the  Fermi surface, it is expected to disappear with doping. 
This is indeed what we found but still, this phase can 
persist over a finite range of doping, similarly to what is found at the mean-field level. 

Using a recently developed analytical continuation technique, we have computed dynamical 
spectral functions. In the SFP phase, we clearly see Dirac cones forming 
around $(\pi/2,\pi/2)$ and equivalent $k$-points, where the dispersion becomes linear. 
Up to a critical doping, these structures are stable so that the Fermi surface
becomes pocket-like.

\begin{acknowledgments}
  We would like to thank C. Wu and S.-C. Zhang for motivating this work and useful discussion. 
Financial support from the 
Bayerisch-Franz\"osisches Hochschulzentrum / Centre de Coop\'eration  Universitaire Franco-Bavarois
is acknowledge. S.~C. thanks IDRIS
(Orsay, France) for use of supercomputer facilities.
\end{acknowledgments}

\end{document}